%% file: E8-FINAL.tex
\font\eightrm=cmr8 at 8pt
\documentclass{oxarticle}
\usepackage{amsmath, amssymb}
\usepackage{graphics, setspace}
\usepackage{epstopdf}
\usepackage{bm}
\usepackage{tikz}
\usepackage{xcolor}

\input{Johnmacros29}

\newcounter{orange} 
\newcounter{apple} 
\addtocounter{apple}{1}
\newcounter{grape} 
\addtocounter{grape}{1}
\newcommand{\numberhere}{E8FINAL }
\newcommand{\articlenumber}{E8-FINAL.tex} 

\usepackage{color}

\newcommand{\mathsym}[1]{{}}
\newcommand{\unicode}[1]{{}}

\usepackage{tikz}

\begin{document}
%\pagestyle{empty}

%\vspace*{1in}
 \begin{center}

{   \huge 
\bf
Removal of the Tachyons from the Fermionic Sector of the Quadratic ZX Action for the Exotic Model 
(E8)
\\
[1cm]}  
%{    J. A. Dixon}
%

\renewcommand{\thefootnote}{\fnsymbol{footnote}}
%\footnotetext[1]{~here we have a footnote.}\renewcommand{\thefootnote}\arabicfootnote}}

{ John A. Dixon\footnote{jadixg@gmail.com, john.dixon@ucalgary.ca}\\Physics Dept\\University of Calgary \\ 
\today 
\\[1cm]}  
%\vfill
 \end{center}
 
 \Large
 
 \Large

  \begin{center} Abstract
 \end{center}

\bf

This paper  contains an encouraging result for the development of the Exotic Model (the `XM'). The XM is made from the \SSM\ (the `SSM')  by coupling it to the `Exotic Invariant'.  This  rather unique  `Exotic Invariant'  is suggested by the detailed form of the SSM, combined with the detailed form of  the BRS cohomology of SUSY. 

 The motivation for examining the XM is that the usual SSM has some well-known, and serious, problems in its generally accepted, but unsatisfactory,  SUSY breaking mechanism.  That mechanism  is based on the spontaneous breaking of SUSY.  

The Exotic Model avoids all of these problems, easily, because its SUSY splitting mechanism does not use, or need, any kind of spontaneous breaking of SUSY. The XM contains a mechanism for  SUSY mass splitting because the addition of the \EI\ modifies  the algebra of SUSY.  This paper is E8 in a series of papers denoted $\rm En, n= 1,2,3\cdots$., all devoted to the XM. 

There is a lot of work still required to elucidate how the XM functions, and to see if the XM is really a viable alternative to the SSM.     The first crucial question, raised in E7, is whether it  is possible to constrain the initial XM action, so that the result excludes tachyonic excitations at tree level.  If the XM is to be a viable theory, it is essential to first remove the tachyonic excitations, in a sensible way, from the fermionic and the bosonic sectors of the quadratic ZX action.

This paper E8 explains how to remove the tachyons   from the basic fermionic `ZX sector' at tree level.   There are four fermion fields in this ZX sector.  The removal requires that all of these fields have the same mass. This result is shown using a Mathematica notebook, which is issued with this paper. 

\tableofcontents

  \section{Introduction}

This is paper E8 in a series of papers  focussed on the `Exotic Model' (the `XM') \ci{E1,E2,E3,E4,E5,E6,E7}. The XM is presented in E6 
\ci{E6}.  In paper E7 \ci{E7} a fundamental and crucial question was raised, and this paper is designed to answer that question, for the fermionic sector. We hope to return to the bosonic sector in E9.

The usual SUSY breaking mechanism for the SSM  has three very serious, and well-known, problems \ci{Allanach:2024suz,D’Onofrio} \ci{weinberg,witten,haberetal,xerxes,bagger}:
\bitem
\item
 There is no set of chiral scalar susy representations that can even approximately account for the experimental results\footnote{All sets seem to have lots of problems--they do not give a spectrum with the superpartners lighter than the observed particles, they do not agree with the observed absence of flavour changing neutral currents etc. These and other issues are all well explained in the literature, for example in the recent reviews \ci{Allanach:2024suz,D’Onofrio} }.   
This problem has given rise to the notion of the `hidden sector' in the literature, which is not satisfactory, since it is more like an admission of failure, and it does not predict much. 
\item
In the usual SSM, the vacuum energy, which becomes the cosmological constant for the related local SUSY theory, is proportional to the VEVs of  the auxiliary fields of the assumed chiral scalar superfields.  This gets a preposterous value in the SSM: it is a google ($10^{100}$) times too large to agree with the experimental result, which is very tiny. This has given rise to the `anthropic argument'. 
\item
 Attempts to mend these problems with `phenomenological soft splitting' terms, added to the SSM,  are also very unconvincing.  These  result in over a hundred unknown and uncalculable parameters, with no explanation of how or why this  SUSY breaking should happen.   
\item
 It is perhaps not surprising that SUSY has not yet been observed, since there is really no sensible theory that predicts how 
the SUSY mass degeneracy can be split, or how SUSY could be found\footnote{There is an argument, called the `anthropic argument', but a difficult problem is whether  there  is any way to test it or disprove it.  The same sort of problem  arises for the `hidden sector'. These questions are discussed in the literature \ci{Allanach:2024suz,D’Onofrio} \ci{weinberg,witten,haberetal,xerxes,bagger}}.
\eitem

\subsection{Brief Summary of the XM}
Here is a brief summary of the present situation for the XM and the E papers.
 The XM is composed of  the \SSM\ plus the special Exotic Invariant\footnote{There is really only one \EI\ which can couple to the SSM, with an unimportant possible addition that makes no difference to the mass splitting.}   described in E6. The \EI s are derived in E2.  The \EI\ arises from the BRS cohomology of supersymmetry, and it necessarily contains a Chiral Dotted Spinor Superfield (a `CDSS').
The XM contains a number of promising features for supersymmetry mass splitting, which is the crucial problem for the SSM.  However it has some worrying features too. This paper discusses the first worrying issues, raised in E7, as we shall describe below.  

\subsection{The Tachyon Problem in the XM: a conjecture that the tachyons can be removed in the full XM, both for fermions and for bosons}

The most obvious of the problems for the XM is that  the CDSS has higher derivatives in its simplest kinetic action, as discussed in E7.  So the XM necessarily contains higher derivative term in its kinetic action, because it contains the CDSS.  These higher derivative terms are associated with the presence of tachyons, which are not acceptable in quantum field theory.

In E7, we noted that the tachyon problem is soluble for the free CDSS theory simply by choosing the ratio of the two possible kinetic terms to be such that the $\Box^2$ terms cancel each other in the propagator. That raised the question whether a similar phenomenon can happen in the full XM, and in particular in the ZX sector at tree level.  This paper E8 confirms that this basic idea also works for the full XM, but it is, understandably, a lot more complicated for the XM than it was for the free CDSS. The \Mn\ that accompanies this paper shows how to remove the tachyons, and it takes approximately 100 seconds to run.

For the free CDSS, the choice of the two kinetic terms removes  the bosonic tachyons as well as the fermionic tachyons,   with the same choice of the two possible kinetic terms.  Supersymmetry is present both before and after gauge symmetry breaking and supersymmetry splitting in the XM, so we conjecture that it is possible to eliminate the tachyons also for the bosons in the XM.  In this paper we show that  the tachyons can be removed  for the fermions in the ZX sector. We hope to return to the bosons of the ZX sector, and to the requirements of SUSY, in E9 and E10.  

As noted in E6 and E7, we anticipate that all the SUSY splitting at tree level in the XM  originates in the ZX sector.  Then the rest of the model, including the quarks, leptons, W bosons, photon, and the Higgs, and all their superpartners, all get SUSY mass splitting from one loop diagrams.

\section{Merits of the XM and its SUSY Splitting Mechanism, compared to the Spontaneous SUSY Breaking Notion in the SSM}

The major merit of the XM, compared with the usual SSM, is of course that we know exactly what the particle content of the XM is. It contains simply the particles of the SSM, plus one new neutral CDSS multiplet.  There is no need for a hypothetical `hidden sector' where SUSY breaking occurs.  There  is no need for a search for a hypothetical set of chiral multiplets that break SUSY spontaneously.  There is no need for an anthropic argument.  All the gauge symmetry breaking arises from the J, H and K chiral multiplets, and the SUSY splitting originates there too.  

So if things continue to work as they do in this E8, we can anticipate that this XM should make some very restrictive and testable experimental predictions.  Those predictions may, of course, be simply wrong.  This paper E8 is a start on making those predictions. 

And, of course, as mentioned in E6 and E7,  the XM accounts in a completely natural way for the observed suppression of flavour changing neutral currents, simply because all the SUSY splitting originates in the neutral ZX sector, which is flavour diagonal.  

Without further ado, we will present the basic result of this paper.
Then we will discuss the problem of renormalization in simple and non-simple BRS cohomology theories.  There is a major unsolved problem there, as will be discussed.   Then we will discuss how this result comes about in the detailed computation of the \Mn\ which accompanies this E8  paper.

\section{Solution of the Tachyon Problem for the fermions of the XM: the result of this  E8 paper}

We start  with a set of dimensionless parameters  $k_i, i = 1 \cdots 7,$ and masses $m_i,  i = 1 \cdots 5$,  for a total of 12 physical parameters.  SUSY is implicit in the form of this action, since it is based on the CDSS and the XM, but we leave room for renormalization effects in these parameters.  This action is essentially the same as the action in equation (73) of section 7 in E7. The notation is changed slightly.  In particular we note that  equation (73) of section 7 in E7 does contain the conterterms analyzed in subsection 9.1 of E7, and these turn out to be essential for the tachyon removal here in E8.  This is probably connected with the puzzle and problems discussed below in section \ref{thenonsimplecase} of this E8 paper. 

To remove the tachyons and higher derivatives from this action, we find that we must impose  a number of fairly simple constraints, which mean that these 12 free coefficients are reduced to 7 free parameters:  three coefficients 
 $k_3, k_5, k_6$ out of the original 7, and  four masses $m_1,m_2,m_3,m_5$ out of the original 5. So four dimensionless coefficients  $k_1, k_2, k_4, k_7$ and one mass $m_4$ 
can be expressed in terms of the remaining coefficients $k_3, k_5, k_6$ and $m_1,m_2,m_3,m_5$. This means that we use up five coefficients to remove the tachyons.

In the accompanying \Mn\  \ci{mnbk} we see that these five constraints  are: 
\be
{\rm firstrule}= 
\left\{k_2\to \frac{2 k_6 m_2}{m_1},k_4\to
   \frac{k_6 m_1}{2 m_2}\right\}
 \la{firstrule}  \ee

\be
{\rm secondrule}=
\left\{k_1\to \frac{2 k_6 m_3^4}{k_5^2 m_1
   m_2^3}-\frac{2 k_3 m_3^2}{k_5
   m_2^2}+\frac{2 k_6 m_2}{m_1}+\frac{k_3^2
   m_1}{2 k_6 m_2},
   \ebp
   k_7\to \frac{k_3 k_5 m_2
   m_1^2-2 k_5 k_6 m_2^2 m_1-2 k_6 m_3^2 m_1+4
   k_6 m_2 m_5^2}{2 k_5 m_1 m_2^2},
   \ebp
 m_4\to \frac{k_3
   k_5 m_1}{2 k_6}-\frac{m_3^2}{m_2}\right\}
 \la{secondrule}   \ee

Note that the right hand sides of these equations consist only of expressions made of the remaining variables $k_3, k_5, k_6$ and $m_1,m_2,m_3,m_5$, and the left hand sides consist of the five parameters that get eliminated\footnote{There are also parameters $e_i, i = 1 \cdots 4$ that we use to label the sources and  masses $M_1,M_2,M_3,M_4,M_5,M_6,M_7$ that we include in the shifts, but these are not physically meaningful.  They are useful for keeping track  of the various terms, however, as is discussed in the \Mn\ \ci{mnbk}.}.

With these choices for the parameters, the propagators reduce to  the `normal form':
\be
 \fr{r_i}{ \Box - m^2}\equiv\fr{r_i}{-\pa_0^2 + \pa_i^2 - m^2} \equiv
 \fr{r_i}{p_0^2 - p_i^2 - m^2}
\la{normalformprop}\ee
where the $r_i, i = 1 \cdots 16$ are dimensionless parameters  made from the dimensionless parameters $k_3, k_5, k_6$   and masses $m_1,m_2,m_3,m_5$  in the theory, and the mass $m^2$ is a simple function, namely 
\be
m^2 =-\frac{k_5 m_1 m_2}{2
   k_6},  
\la{massoftachyfermions}
\ee
 as shown in equation (\ref{valueofmass})  below.

Three of the sixteen $r_i$ are zero. The other 13 non-zero $e_i, i = 1 \cdots 16$  are rather complicated functions of the dimensionless parameters and masses in the theory.

\subsection{The 16 propagators for the fermions reduce to the normal form in the following way:}

\la{sixteennormalprops}

Here we exhibit the solution without tachyons for the fermionic part of the quadratic action. This is what we get if we use  (\ref{firstrule}) and (\ref{secondrule}) on the general solutions.  The parameters $r_i$ in (\ref{normalformprop}), 	and the mass (\ref{massoftachyfermions}) arise, as shown below, from these equations:

\normalsize

\be
{\rm Propagators}=
\left\{\text{prop1}\to 0.\ebp \text{prop5}\to
   -\frac{k_5 m_1^2 m_2 s_1}{\left(2
   \Box k_6+k_5 m_1 m_2\right)
   \left(k_5 m_1 m_2^2-2 m_5^2 m_2+m_1
   m_3^2\right)}.\ebp \text{prop9}\to -\frac{2 k_5
   m_1 m_2^2 s_1}{\left(2 \Box
   k_6+k_5 m_1 m_2\right) \left(k_5 m_1
   m_2^2-2 m_5^2 m_2+m_1
   m_3^2\right)}.\ebp \text{prop13}\to
   0.\ebp \text{prop6}\to -\frac{m_1 m_2 \left(k_5
   m_1 m_2-2 m_5^2\right) s_4}{\left(2
   \Box k_6+k_5 m_1 m_2\right)
   \left(k_5 m_1 m_2^2-2 m_5^2 m_2+m_1
   m_3^2\right) M_5}.\ebp \text{prop8}\to \frac{2
   m_1 m_2 m_3^2 s_4}{\left(2 \Box
   k_6+k_5 m_1 m_2\right) \left(k_5 m_1
   m_2^2-2 m_5^2 m_2+m_1 m_3^2\right)
   M_4}.\ebp \text{prop10}\to -\frac{2 k_6 s_4}{m_1
   m_2 k_5^2+2 \Box k_6
   k_5}.\ebp \text{prop16}\to 0.\ebp \text{prop3}\to
   \frac{2 k_5 m_1 m_2^2 s_3}{\left(2
   \Box k_6+k_5 m_1 m_2\right)
   \left(k_5 m_1 m_2^2-2 m_5^2 m_2+m_1
   m_3^2\right)}.\ebp \text{prop7}\to -\frac{4 m_1
   m_2 \left(k_5^2 k_6 m_2^4-k_3 k_5 m_1 m_3^2
   m_2+k_6 m_3^4\right) s_3}{k_6 \left(2
   \Box k_6+k_5 m_1 m_2\right)
   \left(k_5 m_1 m_2^2-2 m_5^2 m_2+m_1
   m_3^2\right){}^2}.\ebp \text{prop11}\to \frac{2
   m_1 m_2 m_3^2 s_3}{\left(2 \Box
   k_6+k_5 m_1 m_2\right) \left(k_5 m_1
   m_2^2-2 m_5^2 m_2+m_1 m_3^2\right)
   M_1}.\ebp \text{prop15}\to -\frac{m_1^2 \left(2
   k_6 \left(k_5^2 m_2^4+m_3^4\right)+k_3 k_5
   m_2 \left(k_5 m_1 m_2^2-2 m_5^2 m_2-m_1
   m_3^2\right)\right) s_3}{k_6 \left(2
   \Box k_6+k_5 m_1 m_2\right)
   \left(k_5 m_1 m_2^2-2 m_5^2 m_2+m_1
   m_3^2\right){}^2}.\ebp \text{prop2}\to
   -\frac{k_5 m_1^2 m_2 s_2}{\left(2
   \Box k_6+k_5 m_1 m_2\right)
   \left(k_5 m_1 m_2^2-2 m_5^2 m_2+m_1
   m_3^2\right)}.\ebp \text{prop4}\to 
   \ebp
   \frac{m_1
   \left(k_3 k_5 m_2 \left(2 m_5^2-k_5 m_1
   m_2\right) m_1^2+2 k_6 \left(2 m_2 m_5^4-2
   m_1 m_3^2 m_5^2+k_5 m_1 m_2 \left(m_1
   m_3^2-2 m_2 m_5^2\right)\right)\right)
   s_2}{k_6 \left(2 \Box k_6+k_5 m_1
   m_2\right) \left(k_5 m_1 m_2^2-2 m_5^2
   m_2+m_1 m_3^2\right){}^2}.\ebp \text{prop12}\to
   -\frac{m_1 m_2 \left(k_5 m_1 m_2-2
   m_5^2\right) s_2}{\left(2 \Box
   k_6+k_5 m_1 m_2\right) \left(k_5 m_1
   m_2^2-2 m_5^2 m_2+m_1 m_3^2\right)
   M_2}.\ebp \text{prop14}\to -\frac{m_1^2 \left(2
   k_6 \left(k_5^2 m_2^4+m_3^4\right)+k_3 k_5
   m_2 \left(k_5 m_1 m_2^2-2 m_5^2 m_2-m_1
   m_3^2\right)\right) s_2}{k_6 \left(2
   \Box k_6+k_5 m_1 m_2\right)
   \left(k_5 m_1 m_2^2-2 m_5^2 m_2+m_1
   m_3^2\right){}^2}\right\}\ee

\Large

The following denominator occurs in every one of the above: \be
 \left(2
   \Box k_6+k_5 m_1 m_2\right) 
  = denom
\ee   
and it corresponds to the same mass for each non-zero propagator:
\be \left\{\Box\to
   -\frac{k_5 m_1 m_2}{2
   k_6}\right\} 
   \la{valueofmass}
\ee

\section{The parameters, SUSY and the question of Renormalization of the BRS transformations in the XM}

There are many questions that arise now.  We do not claim to include SUSY very carefully at this point in this calculation, and we have been rather vague about how to write down the action that we use in the \Mn\ \ci{mnbk}, and in the sections above.  These are indeed problems, and the next sections are meant to explain what needs to be done to deal with these problems.

These results will be discussed again below,  after we have discussed some important features of the BRS cohomology and SUSY.

 \section{The Master Equation for BRS symmetry,   the old conjecture for the `simple case',  and the new `non-simple case'}

In this section, we will review the construction of the Renormalized  Action for the `simple case',  where the BRS cohomology is `simple', and also for the `non-simple case', which is what arises for SUSY:  
\ben
\item
By `simple' here we mean the case where the `old conjecture' is true\footnote{ The point of the XM is that the old conjecture is not true for rigid SUSY.  But here we will first pursue the old ideas, and later return to the new situation for the XM and the ZX sector. This author has never been satisfied with the `proofs' of the old conjecture that have been offered by various authors. This is the reason that the paper \ci{E2} was written.  It shows that the old conjecture is not true for SUSY.  The spectral sequence method shows this, and as far as this author knows, that method has not been used in work by other authors for BRS cohomology.}, By `non-simple' we mean the case where the `old conjecture' is not true. 
\item
Here is  the `old conjecture' for a given theory:
\begin{verse}
The old conjecture is that every object in the BRS cohomology space of the theory can be written so that it contains only fields, without any pseudofields or ghosts. This is believed to include Yang-Mills theory without SUSY, but this author is not satisfied that there is a   proof that it is true even for that case\footnote{This is a complicated subject.  This author thinks that only the spectral sequence can generate a proof that is believable, and that has not yet been done.  Presumably a proof could be constructed along he lines of E2, but that has not yet been done.}.  
This is called the `simple case'.
\end{verse}
\item
We now need to explain the non-simple case for a given theory:
\begin{verse}
The non-simple case is that there are some objects in the BRS cohomology space of the theory which can only be written  in a way that includes pseudofields and ghosts. 
This includes rigid SUSY, as was shown in E2. This is called the `non-simple case'.

\end{verse}
\item
Now we must explain what the BRS cohomology space is:
\ben
\item
For any quantum field theory with a relevant nilpotent BRS variation one can construct a Master Equation, which, for the present case,  has the form 

\be
\cM= \int d^4 x \lt \{ \fr{\d \cA_{\rm Total}}{\d a }\fr{\d \cA_{\rm Total}}{\d {\widetilde  a}}+ \fr{\d \cA_{\rm Total}}{\d \y_{\a} }\fr{\d \cA_{\rm Total}}{\pa {\widetilde  \y}^{\a}}+ \cdots \rt \}
\ee
where the action contains terms of the form
\be
\cA_{\rm \Psf s} = \int d^4 x \lt \{{\widetilde a} \d a +{\widetilde  \y}^{\a}\d \y_{\a} +\cdots \rt \}
\la{AZinn}
\ee
where the BRS transformation $\d$ is Grassmann odd and also nilpotent
\be
\d^2 = 0
\ee
and  the initial unrenormalized action has the form 
\be
\cA_{\rm Total} = \cA_{\rm Zinn} +\cA_{\rm Invariant}
\la{Atotal}
\ee
where
\be
\cA_{\rm Invariant} = \int d^4 x \lt \{ a \Box \oa + \y \pa \oy  - \fr{1}{4} F^{\m\n} F_{\m\n} + \cdots \rt \}
\la{AInv}
\ee
is a sum of invariants, which are functions only of the fields, and not of the pseudofields or ghosts\footnote{Remember that for Yang-Mills theory, and gravity, and supergravity, it is necessary to add a gauge fixing and ghost action to the theory. This can be expressed as a boundary of the BRS operator.},  which satisfy
\be
\d \cA_{\rm Invariant} =0
\ee
and we also require that each separate invariant in $\cA_{\rm Invariant}$ be in the BRS cohomology space of $\d$.  This means that there exist no $\cA_{\rm Boundary}$ such that 
\be
 \cA_{\rm Invariant}=\d \cA_{\rm Boundary}
\ee 
where $\cA_{\rm Boundary}$ has the following characteristics
\ben
\item
$\cA_{\rm Boundary}$ is the integral of a local functional of the fields pseudofields, ghosts and the derivative $\pa_{\a \dot \b}$
\item
$\cA_{\rm Boundary}$ must have ghost charge minus one, where the  ghost charge is defined so that $\d$ increases ghost charge by one, and all the pseudofields have  ghost charge defined so that ghost charge is conserved.

\een
\item
In this case one can express the renormalized action as a sum of invariants plus a 
Canonical Transformation as described in \ci{Dixon:1975si}.
\een

\een

 \subsection{The Non-Simple Case for the BRS Cohomology}
\la{thenonsimplecase}
When we come to the non-simple case, the situation becomes complicated and  challenging. The usual methods \ci{quarkmasses,mcleary,Kluberg-Stern:1974iel,Kluberg-Stern:1975ebk,Kluberg-Stern:1974nmx,Dixon:1974ss,Dixon:1975si,Deans:1978wn} \ci{holes,holescommun,dixminram} are not easy to apply here.   Remember that for SUSY we also have a constraint equation \ci{E2}, in addition to items in the cohomology space that depend crucially on the pseudofields and the ghosts. 
The constraint equation is discussed in E2 and E6 and E7 \ci{E2,E5,E7}.  It is a crucial feature of the XM.  

 The consequences of this combination need to be understood in connection with the counterterms that arise upon renormalization at one (or several) loops.  That task will not be attempted here.  It clearly needs to be understood, but for the present paper we will examine the propagators without understanding these issues. 
The results of this paper are already rather complicated anyway.  Do they relate to this question?  It seems very likely that they do.  That is an unsolved problem.  It needs attention.

\subsection{Comparison of the \Mn\ and the Results here}

In E7, we wrote out the problem for the fermion propagator of the ZX sector in subsection 7.1.  The assignments of the parameters $a_i$ and $m_i$ there and $k_i$ and $m_i$ in the \Mn\ are a bit different.  Here are the important features however:
\ben
\item
In E7 it was not clear whether we could simply proceed to use the unrenormalized action.  Now it is clear that we should use the renormalized action.  The unrenormalized action will not give us any solution for the tachyon problem.  The solution for the tachyon problem in this E8 comes about precisely because we include the terms that arise from renormalization.  
\item
This is connected with the puzzles in subsection \ref{thenonsimplecase} above. 
\item It is hoped that these questions will become clearer when we examine the ZX bosons and consider the renormalization of SUSY that takes place for this non-simple BRS cohomology.
\een

\section{The Fermions in the ZX sector}

Our Lagrangian for the fermions in the \Mn\  \ci{mnbk} is named ``quadfermionlagrangian" there.  It is essentially the same as the action that is represented in the expression in subsection 7.1 in E7, though the notation is a bit different.  

The general solution for the propagators $\rm \D_n\equiv propn$ in the \Mn\ in \ci{mnbk}, as repeated above in subsection (\ref{sixteennormalprops}), has the following features:
\ben
\item
it says that all four of the fermions must have the same mass m,
\item
it says that there are no tachyons present, because there are no higher derivatives, 
\item
it generates quite complicated functions for the dimensionless factors $r_i$.  
\een

It is not easy (at least for this author) to understand how or why the theory generates these  results for the propagators.   There must  be a good and simple reason for this, but that reason is not evident, and since it arises in a very complicated way, after thousands of terms, and 100 seconds of computation in the  \Mn\ \ci{mnbk}, it is hard to see why it happens. The result does seem to be consistent with the ideas that nature abhors a tachyon and that the BRS SUSY cohomology contains some physics, but those are not satisfactory explanations, of course.

\section{Discussion of the \Ma\ Notebook in \ci{mnbk} and the  Tachyon Problem }

In E7, we made  a reasonable guess for the renormalized ZX quadratic action.  The next problem was to attack the question of the spectrum of the model by constructing the propagators.  The spectrum can be deduced from them.  In this section we will go through the steps that take place in  the \Ma\ Notebook in 
\ci{mnbk}.

The propagators are generated as follows. We need to use somewhat different notation in \Ma, and so we must juggle back and forth for the notation:
\ben
\item
First we add the following source terms to the action
\[
\cL_{\rm Source} =   \lt \{  e_1
{\widetilde \y}^{\a} \y_{\a} +e_2
{\widetilde \lam}^{\a} \lam_{\a} +e_3
{\widetilde \c}^{\dot\a} \c_{\dot\a} +e_4
{\widetilde \f}^{\dot\a} \f_{\dot\a} \rt. \] \be \lt.
+\rm \CC
\rt \}
\equiv
   \lt \{  e_m
  {\rm Source}_m     {\rm Field}_m 
\rt \}
\la{Asource} 
\ee 
This idea comes from the method used to find propagators in the path integral formalism, where one introduces shifts and then insists that the integral over the fields is just a constant, leaving a generator for the propagators.  This happens if the `mixed terms' are set to zero, as seen in the \Mn \ci{mnbk}.
\item
The next step is to introduce the shifts and the shifted fields.  The first of sixteen of these has the following form for this fermion part:
\[
 \psi_{\rm 0 \;shifted}=\psi_0\to
\psi_0
   -\fr{ \pa_{ 0,1 } \D_{13} {\widetilde\phi}_0}{M_3}
   +\fr{ \pa_{ 0,0 } \D_{13} {\widetilde\phi}_1}{M_3}
\] \be  
  +  \pa_{ 0,0 } \D_{10} {\widetilde{\ov\psi}}_1 
   - \pa_{ 0,1 } \D_{10} {\widetilde{\ov\psi}}_0 
+i M_1
   \D_{11} {\widetilde\lambda}_0  +M_2
\D_{12} {\widetilde\chi}_0  
\ee
and this takes the following form in the notation of the \Mn:
\be
 \left. \text{$\psi $shift0}=\psi
   _0\to 
+\psi_0
   -\frac{\text{pa}_{\{0,1\}}.\text{prop13}.\text{sour$\phi$}_0}{M_3}
\ebp
+\frac{\text{pa}_{\{0,0\}}.\text{prop13}.\text{sour$\phi
   $}_1}{M_3}
+\text{pa}_{\{0,0\}}.\text{prop10}.\text{sour$\psi
   $bar}_1
\ebp
-\text{pa}_{\{0,1\}}.\text{prop10}.\text{sour$\psi $bar}_0+i M_1
   \text{prop11}.\text{sour$\lambda $}_0
\ebp+M_2
   \text{prop12}.\text{sour$\chi $bar}_0\right.
\ee
\item
The form of these shifts is obtained as follows:
\ben
 \item
First we look at the action and note that from the kinetic and mass terms in it, we can obtain epressions that look like (\ref{Asource}) if we add terms as shown, provided that the $\D_n$ are suitable.  
\item
For the fermions, we add four source terms to each fermion for the shift.  This is a general shift. 
Since there are 4 different fermion fields, (each with two spin types, and each with a complex conjugate), we need 16 $\D_n$ to implement these shifts.  We can assume that the $\D_n$ are real, so the \CC s are determined easily.
\een
\item
The next step is to introduce the shifts into the quadratic renormalized action and put that action into a canonical form of some kind.  That is the major task of the \Mn. Then we collect all the terms which are of the form
\be
\cL_{\rm Linear\;in\;Souces} = \S_{l,m,n,p,q,r} \ 
 f_l[e,k]  {\rm Source}_m  (\pa)^n M^p   \D_{q}   {\rm Field}_r = 0
  \ee
and we have set all these terms to zero to generate equations, and we solve for the $  \D_{q} $ as a function of the  $e_r, k_l ,  (\pa) , M  $
, after eliminating the ${\rm Source}_m $ and the ${\rm Field}_r $ to generate a set of linear equations. 
\item
Typically we get something like
\be
 \D_{q} \approx \fr{ \Box^2 +  \Box m^2 +m^4  }{ \Box^3 + \Box^2 m^2 +\Box m^4 + m^6  }
\ee
where the coefficients   $e_r, k_l,m_i$ are left implicit. If we can find constraints of the form
\be
k_i \ra f_{\rm i, Constraint} [k,e]
\la{ConstraintsforNormal}
\ee
 on the coefficients $e_r, k_l,m_i$ such that the all propagators in the above reduce to ($g_q$ and $s_q$ below are dimensionless)
\be
 \D_{\rm Normal, q}  \ra  \fr{ g_q[k,e,m] }{ \Box -  s_q[k,e,m]  m^2  }
\la{Normalprops}
\ee
then we will say that the constraints reduce the propagators to normal form.
\item
Full removal of the tachyons at tree level requires normal form plus some restrictions about signs. 
\item 
The next step would be  to express the propagators in the form with two sources, which would result in something like the following genertor for the propagators in the path integral formalism:
\be
\cG=\S \int \int \lt \{
{\rm  source}. \D . \pa .{\rm  source} \rt \}
\ee
but we will not attempt to write that down here.

\een

\section{More Discussion }

\la{explanationofresults}

The crucial question raised in E7 was whether it was possible to get propagators  that do not include tachyons.     This question is answered in the affirmative in this E8 paper.

In E7 it was shown that, for the free theory, the CDSS can be written without tachyons by choosing a certain combination of the two possible kinetic terms there.  That suggests that it might be possible to do the same for the full interacting theory.  It turns out that this actually can be done, but there are plenty of mysteries still here.

\section{The Structure of the Result for the Fermions}

In section (\ref{sixteennormalprops})  we have 13 healthy looking propagators with the same mass and 3 that are zero.  What does this mean? The author admits that he does not understand why this happens.  The reason why this happens is not easy to see since the \Ma\ notebook \ci{mnbk}  is quite complicated, although the answer is simple, in a way.

 In\Ma\ notebook \ci{mnbk}  we  set out the way to solve for the fermionic propagators.   It turns out that the way to solve for the fermionic propagators involves several steps.
\ben
\item
In the first place, the author now believes that it is necessary to include the counterterms in the starting Lagrangian. If this is not done, it is not possible to remove the tachyons.  If this is  done, it is   possible to remove the tachyons.  
\item
Then there are 16 propagators $\Delta_n, n=1\cdots 16$. They are chosen to have dimension -2.  
\item
There are two sets of 8 equations each that relate the propagators $\Delta_n, n=1\cdots 16$ to each other.  
\een 
\section{The Four Quartets}

Examination of those 16 equations shows that they divide into four quartets.    
\ben
\item
The four quartets come from 16 equations, as follows:
\ben
\item ${\rm First}: \Delta_1,\Delta_5,\Delta_9,\Delta_{13}$;
 \item${\rm Second}:\Delta_{6},\Delta_{8}
,\Delta_{10},\Delta_{16}$;
\item ${\rm Third}:\Delta_3,\Delta_7,\Delta_{11},\Delta_{15}$;
 \item${\rm Fourth}:\Delta_2,\Delta_4,\Delta_{12},\Delta_{14}$.
\een
\item
It turns out that all four quartets yield solutions with  the same Denominator. So there are 16 Denominators that are the same in these four quartets.  This Denominator takes the complicated form in section (\ref{initialdenominatorsection}).  We can write it schematically in the form \be
{\rm Denominator}\approx \Box^3 + \Box^2 m^2+ \Box m^4 +   m^6\ee 
\een

\section{The Initial Denominator}
\la{initialdenominatorsection}

Here is the rather formidable denominator which appears in the propagators that we obtain from the starting Lagrangian here.  
\be
  {\rm Denominator}=
4 \Box
   k_4 k_5^3 m_2^4-k_5^4 m_1^2 m_2^4+4 k_5^3
   m_1 m_5^2 m_2^3\eb
-4\Box k_5^3 k_6
   m_1 m_2^3+4\Box k_5^3 k_7 m_1
   m_2^3-4 k_5^2 m_5^4 m_2^2
\eb
+4
   \Box^2  k_1 k_4 k_5^2 m_2^2+4
   \Box^2  k_2 k_4 k_5^2 m_2^2-4
   \Box^2  k_5^2 k_6^2 m_2^2
\eb
-4
   \Box^2  k_5^2 k_7^2 m_2^2-2
  \Box k_3 k_5^3 m_1^2 m_2^2-2
   k_5^3 m_1^2 m_3^2 m_2^2-8 \Box
   k_5^2 k_7 m_5^2 m_2^2
\eb
-4 \Box
   k_5^2 k_6 m_1 m_4 m_2^2-4 \Box
   k_5^2 k_6 m_1 m_3^2 m_2+4 \Box
   k_5^2 k_7 m_1 m_3^2 m_2
\eb
+4 k_5^2 m_1 m_3^2
   m_5^2 m_2+4\Box k_2 k_5^2 m_1
   m_5^2 m_2
\eb
T+4\Box k_3 k_5^2 m_1
   m_5^2 m_2+8\Box k_5 k_6 m_4
   m_5^2 m_2
\eb-4 \Box^2  k_1 k_5^2 k_6
   m_1 m_2-4 \Box^2  k_3 k_5^2 k_6
   m_1 m_2
\eb+4 \Box^2  k_2 k_5^2 k_7
   m_1 m_2+4 \Box^2  k_3 k_5^2 k_7
   m_1 m_2\eb
+2\Box k_2 k_5^2 m_1^2
   m_4 m_2-8\Box k_4 k_5 m_3^2 m_4
   m_2
\eb
-8 \Box^2  k_3 k_4 k_5 m_4
   m_2+8 \Box^2  k_5 k_6 k_7 m_4
   m_2-k_5^2 m_1^2 m_3^4
\eb
-4\Box k_4
   k_5 m_3^4-4\Box k_2 k_5 m_5^4-4
   \Box^3 k_1 k_5 k_6^2-4
   \Box^3 k_2 k_5
   k_7^2-\Box^2  k_3^2 k_5^2
   m_1^2
\eb
+\Box^2  k_1 k_2 k_5^2
   m_1^2-2\Box k_3 k_5^2 m_1^2
   m_3^2-8 \Box^2  k_3 k_4 k_5
   m_3^2+8 \Box^2  k_5 k_6 k_7
   m_3^2
\eb
-4 \Box^2  k_6^2 m_4^2+4
   \Box^2  k_2 k_4 m_4^2+8
  \Box k_5 k_6 m_3^2 m_5^2
\eb
+8
   \Box^2  k_3 k_5 k_6 m_5^2-8
   \Box^2  k_2 k_5 k_7 m_5^2-4
  \Box k_2 k_5 m_1 m_4 m_5^2-4
   \Box^3 k_3^2 k_4 k_5
\eb
+4
   \Box^3 k_1 k_2 k_4 k_5+8
   \Box^3 k_3 k_5 k_6 k_7+4
  \Box k_5 k_6 m_1 m_3^2 m_4
\eb
+4
   \Box^2  k_3 k_5 k_6 m_1 m_4-4
   \Box^2  k_2 k_5 k_7 m_1 m_4
\ee
Note that it contains  $\Box^3$ and $\Box^2$ terms in addition to $\Box$ and mass terms.
As mentioned above, we can write this schematically as
\be
 {\rm Denominator}  \approx \Box^3 + \Box^2 m^2+ \Box m^4 +   m^6
\ee
where we ignore all the details.

Application of the `firstrule' in (\ref{firstrule}) makes this factorize into three `normal' terms as follows:
\[
 {\rm Denominator} \ra k_5 \left(2 \Box k_6+k_5 m_1
   m_2\right) 
\]\be
\frac{
\left(-2  \Box k_7
   m_2+ \Box k_3 m_1+k_5 m_1 m_2^2-2
   m_5^2 m_2+m_1 m_3^2\right){}^2}{m_1 m_2}
\la{factorsfordenom}
\ee
That is the first step which makes the tachyons disappear here.  Why does it happen?  

\section{A Slightly Philosophical Approach?}
\la{philo}

  The general solution for the $\Delta_n$ can be written in the form ``solsfrom4quartets'' in   \ci{mnbk}. The question then was this: ``How can one find a way to convert this formidable set of formulae to a set of 16 healthy looking propagators?''  The technique used was  to assume that there is a way to do it.  That leads to the following method.  By inspection of the solutions ``solsfrom4quartets'', we can see that thirteen of these sixteen  propagators $\rm propn, n = 1 \cdots 16$  have   the following schematic form
\be
{\rm propn \ra \fr{     \Box^2  + \Box m^2 +   m^4}{  \Box^3 + \Box^2 m^2+ \Box m^4 +   m^6}}
\la{thirteennumerators}
\ee
and three of them  have the following schematic form
 \be
{\rm propn \ra \fr{\Box m^2 +   m^4}{  \Box^3 + \Box^2 m^2+ \Box m^4 +   m^6}}
\la{threenumerators}
\ee
 It is clear that propagators  of the form (\ref{threenumerators}) cannot possibly cancel to yield propagators of the normal form in (\ref{normalformprop}), because the powers of $\Box$ do not match.
If (\ref{threenumerators}) are to reduce to the normal schematic form 
\be
{\rm propn \ra \fr{    1}{   \Box + m^2}}
\la{normalform}\ee
then we need to constrain the coefficients so that the three terms  (\ref{threenumerators}) have numerators such that
\be
\Box m^2 +   m^4\ra 0
\ee
This is what gives rise to (\ref{firstrule}).  It makes all three of the numerators with schematic form (\ref{threenumerators}) go to zero, and, at the same time it makes five more numerators go to the form 
(\ref{normalform}).  It leaves 
 8 numerators still problematic.  However the rule (\ref{firstrule}) also causes the denominator in section \ref{initialdenominatorsection} to become factorizable into  the product (\ref{factorsfordenom}), which has the
schematic form 
\be
denom=  (\Box + m_1^2) (\Box + m_2^2)^2 
\la{factordenom}
\ee
Since this is the initial denominator for all the propn, this factorizes for all the propn.
Inspection then shows that are  four terms of the form 
\be
{\rm propn \ra \fr{ k \Box  +   m_3^2}{ (\Box + m_1^2) (\Box + m_2^2) }}
\la{easy}
\ee
and four more terms of the form
\be
{\rm propn \ra \fr{     \Box^2  + \Box m^2 +   m^4}{ (\Box + m_1^2) (\Box + m_2^2)^2}}
\la{hard}
\ee
  
  Then one has several choices as to how to try to reduce these remaining 8 propn to normal form.  The method used was  to try to find a solution so that the terms (\ref{easy}) go to normal form:
\be
{\rm (   k \Box  +   m_3^2)= k_{new,1}   (  \Box  +   m_2^2)}
\la{trick}
\ee
where $k_{new,1}$ is a new dimensionless coefficient.  This works, 
and one can proceed to use the same idea for any remaining problems.  This is how the solution above was found.  There are quite a few possibilities of this kind, but for the purpose of this paper, we are not going to try to examine them all.  It is far from clear why this works, and also it is not clear what other solutions look like. 

Now we know that there is at least one way to do eliminate the tachyons for the fermions in the ZX sector, and it is in subsection (\ref{sixteennormalprops}).

\section{Conclusion}

The main result of this paper E8 is that the forms in subsection  (\ref{sixteennormalprops}) do exist, for this case.  It is not obvious (at least not to this author) that they must exist. Nor is it clear yet whether they are unique in some sense.  Nor is it clear how to describe the relation between the bare action and the renormalized action for this non-simple case of the BRS cohomology.  So there is much work to be done.

If such forms do exist for the bosons too, then that is a significant feature for the ZX sector and for the XM.  There is no point in disussing that until we see whether such solutions also exist for the bosons of the ZX sector.

 In this paper we exhibit an action for the XM which results in a theory without tachyons for the  ZX fermions. All that is required is a special  restriction on the parameters.   The next step is to see whether this can also be done for the bosons. We hope to return to that problem in E9.  

As noted in section \ref{philo}, there may be several ways to do this, or none at all.  

If it can be done for both the fermions and the bosons, using generic forms of the action as was done here for the fermions, the next question is whether one can find a set which is also consistent with SUSY.  As we noted above, this requires a careful analysis of the renormalization of the nilpotent SUSY operator, and of the counterterms that arise from that renomalization. We hope to return to that problem in E9 or E10.

If this program  can be completed without new serious problems, and if the Z boson has a different (smaller) mass from the other bosons (and the fermions), in the ZX sector, then the XM may become a serious contender for an improved, and much more predictive, SSM.  But obviously there are many issues that must be confronted, even if this first tachyon problem is solvable  in the bosonic part of the ZX sector at tree level.   But clearly that is the first problem that now needs solution.

\begin{center}
 { Acknowledgments}
\end{center}
\vspace{.1cm}

  I thank     Friedemann Brandt, Philip Candelas,   Mike Duff, Sergio Ferrara,  Dylan Harries, Marc Henneaux,  D.R.T. Jones, Olivier Piguet, Antoine van Proeyen,     Peter West and Ed Witten for stimulating correspondence and conversations.   I also express appreciation for help in the past from William Deans, Lochlainn O'Raifeartaigh, Graham Ross, Raymond Stora, Steven Weinberg, Julius Wess and Bruno Zumino. They are not replaceable and they are missed.  I also  thank Ben 
Allanach, Doug Baxter,  Margaret Blair,  Murray Campbell, David Cornwell, Thom Curtright, James Dodd, Richard Golding, Chris T.  Hill,  Davide Rovere,   Pierre Ramond, Peter Scharbach,  Mahdi Shamsei, Sean Stotyn, Xerxes Tata and J.C. Taylor, for recent, and helpful, encouragement to carry on with this work. I also express appreciation to Dylan Harries and  to Will, Dave and Peter Dixon, Vanessa McAdam and Sarunas Verner for encouraging and teaching me to use coding.  I note with sadness the recent passing of Carlo Becchi and Kelly Stelle, both of whom were  valuable colleagues and good friends.

 \tiny 
\articlenumber\\
Dec 8,2025

\Large

\end{document}

%% file: Johnmacros29.tex
\newcommand{\Mn}{{\Ma\; Notebook}}

\newcommand{\XM}{{Exotic Model}}

\newcommand{\cM}{{\cal M}}

\newcommand{\CDSS}{{Chiral Dotted Spinor Superfield}}

\newcommand{\Ma}{{Mathematica}}

\newcommand{\cG}{{\cal G}}

\newcommand{\cA}{{\cal A}}

\newcommand{\CC}{Complex Conjugate}

\newcommand{\Exists}{\bm{\exists}\kern-0.6em\bm{\exists}}

\newcommand{\EI}{Exotic Invariant}

\newcommand{\Psf}{Pseudofield}

\newcommand{\SSM}{Supersymmetric Standard Model}

\newcommand{\cL}{{\cal L}}
\newcommand{\bt}{\begin{tabular}{c}}
\newcommand{\et}{\end{tabular}}

\newcommand{\eb}{\ee\be } 
\newcommand{\ebp}{\rt.\ee\be\lt.}

\newcommand{\bmat}{\lt ( \begin{array} }
\newcommand{\emat}{  \end{array} \rt )}

\newcommand{\oa}{{\ov a}}

\newcommand{\oy}{{\ov \y}}

\renewcommand{\a}{\alpha}	
\renewcommand{\b}{\beta}

\renewcommand{\d}{\delta}

\newcommand{\lam}{\lambda}
\newcommand{\m}{\mu}
\newcommand{\n}{\nu}

\newcommand{\f}{\phi}
\renewcommand{\c}{\chi}
\newcommand{\y}{\psi}

\newcommand{\D}{\Delta}

\renewcommand{\S}{\Sigma}

\newcommand{\la}{\label}
\newcommand{\ci}{\cite}

\newcommand{\ds}{\documentstyle}	
\newcommand{\fr}{\frac}

\newcommand{\pa}{\partial}
\newcommand{\ov}{\overline}
\newcommand{\br}{\begin{rant}}
\newcommand{\er}{\end{rant}}
\newcommand{\beC}{\begin{Conjecture}}
\newcommand{\eeC}{\end{Conjecture}}
\newcommand{\be}{\begin{equation}}
\newcommand{\ee}{\end{equation}}
\newcommand{\ba}{\begin{array}} 
\newcommand{\ea}{\end{array}}
\newcommand{\bea}{\begin{eqnarray}}
\newcommand{\eea}{\end{eqnarray}}
\newcommand{\ra}{\rightarrow}

\newcommand{\lt}{\left}
\newcommand{\rt}{\right}

\newcommand{\ben}{\begin{enumerate}}
\newcommand{\een}{\end{enumerate}}
\newcommand{\bitem}{\begin{itemize}}
\newcommand{\eitem}{\end{itemize}}

%% file: E8-FINAL.bbl
\begin{thebibliography}{99}
\bibitem{mnbk} \Ma\ notebook: C: Users Jadix Desktop E8Final E8-Dec8Final.nb,  included with this paper in both pdf and nb forms. More sophisticated coding can be found in \ci{SARAH}, but SARAH does not include  models like the XM of the E series.

 \vspace{.8cm}
Recent reviews about supersymmetry and experiment
\bibitem{E8} Ibid. ``Removal of the Tachyons from the Fermionic Sector of the Quadratic ZX Action for the Exotic Model 
(E8)"    This is the present Preprint,  issued in December, 2025.
\end{thebibliography}
